\documentclass[aps,prl,floatfix,twocolumn,showpacs,reprint]{revtex4-1}
\usepackage{amsfonts}
\usepackage{mathrsfs}
\usepackage{amsmath}
\usepackage{color}
\usepackage{graphicx}
\usepackage{bm}
\usepackage{amssymb}
\usepackage{xspace}
\usepackage{epstopdf}
\usepackage{dcolumn}
\usepackage{longtable}
\usepackage{subfigure}
\usepackage{multirow}
\usepackage[colorlinks=true, letterpaper=true, pdfstartview=FitV, linkcolor=blue, citecolor=blue, urlcolor=blue]{hyperref}

\newcommand{\sgn}{\text{sgn}}

\makeatletter

\newcommand{\Rmnum}[1]{\expandafter\@slowromancap\romannumeral #1@}
\makeatother

\begin{document}
\title{
Topological Mirror Superconductivity
       }
\author{Fan Zhang}\email{zhf@sas.upenn.edu}
\author{C. L. Kane}
\author{E. J. Mele}
\affiliation{Department of Physics and Astronomy, University of Pennsylvania, Philadelphia, PA 19104, USA}
\begin{abstract}
We demonstrate the existence of topological superconductors (SC) protected by mirror and time reversal (TR) symmetries.
$D$-dimensional ($D$$=$$1,2,3$) crystalline SCs are characterized by $2^{D-1}$ independent integer topological invariants,
which take the form of mirror Berry phases.
These invariants determine the distribution of Majorana modes on a mirror symmetric boundary.
The parity of total mirror Berry phase is the $\mathbb{Z}_2$ index of a class DIII SC,
implying that a DIII topological SC with a mirror line must also be a topological mirror SC but not vice versa,
and that a DIII SC with a mirror plane is always TR trivial but can be mirror topological.
We introduce representative models and suggest experimental signatures in feasible systems.
Advances in quantum computing, the case for nodal SCs, 
the case for class D, and topological SCs protected by rotational symmetries are pointed out.
\end{abstract}
\maketitle

The advent of topological insulators protected by time reversal (TR) symmetry~\cite{RMP1,RMP2} opened the door
to the search for other topological states with different symmetries~\cite{Ryu,Kitaev,TMI,TMI-exp1,TMI-exp2,TMI-exp3,Bernevig,Xiao-R,Fiete}.
The idea replacing the winding number of $p$ wave pairing~\cite{pSC1,pSC2,pSC3} by the Berry phase of a single helical band
further provided a promising route to engineer topological superconductivity
using an ordinary superconductor (SC)~\cite{Kane-D,Chuanwei,Fujimoto-D,Sau-D,Alicea-D,Oreg-D}.
The hallmark of these class D topological states, governed by TR symmetry breaking and particle hole (PH) redundancy,
is the existence of Majorana modes on the boundaries.  There is presently a major effort to detect their unique signatures~\cite{Kouwenhoven,Shtrikman,Furdyna,Xu,Brinkman,DGG,FJE,Law1,Beenakker,Wieder,Zhang-DIII}
such as resonant Andreev reflection and fractional Josephson effect.
Recently, realizations of TR invariant topological SCs have also been proposed~\cite{Zhang-DIII,Law2,Volovik,Qi,Fu-DIII,Ando-DIII,Viola,Nagaosa}
in a variety of settings. One might wonder whether a crystalline symmetry can also lead to a different class of topological SCs.

Remarkably enough, there exists an even richer class of topological SCs protected by {\it mirror} and TR symmetries,
as we will demonstrate in this letter.
$D$-dimensional ($D$$=$$1,2,3$) crystalline SCs in this symmetry class are characterized by $2^{D-1}$ integer invariants
determined by the mirror Berry phases of the negative energy bands along the mirror and TR invariant lines.
These determine the distribution of Majorana modes on a mirror symmetric boundary.
Interestingly, the parity of total mirror Berry phase is the $\mathbb{Z}_2$ index of a class DIII SC,
{i.e.}, a fully gapped SC respecting TR symmetry.
This relation leads to two important implications: in 1D and 2D a DIII topological SC with a mirror line
must also be a topological mirror SC but not vice versa;
in 2D and 3D a DIII SC with a mirror plane is always trivial in class DIII (TR trivial)
but can be topological in the class with mirror symmetry (mirror topological).

Now we shall develop a simple symmetry argument, without any calculation,
to demonstrate the stability of $N$ Majorana Kramers pairs (MKP)~\cite{MKP} at a mirror symmetric end.
First consider a 1D DIII topological SC, which exhibits a single MKP at the end~\cite{Zhang-DIII,Law2}.
In the two dimensional subspace spanned by the two Majorana modes, we may choose a gauge in which the antiunitary TR
and PH symmetry operators are $\Theta=\sigma_y K$ and $\Xi=\sigma_x K$, respectively,
with $K$ the complex conjugation and ${\bm\sigma}$ the Pauli matrices.
Now consider $N$ such SCs that physically coincide and each is a mirror line.
This mirror symmetry must be described in this space by $\mathcal{M}=-i\sigma_z$ (up to a sign),
since it squares to $-1$ and commutes with $\Theta$ and $\Xi$.
$\mathcal{H}_M$, the mirror symmetric couplings among the $N$ MKPs, must satisfy the constraints of mirror, TR, and PH symmetries:
$[\mathcal{H}_M,\mathcal{M}]=\{\mathcal{H}_M,\Pi\}=0$,
where $\Pi\equiv\Xi\Theta=-i\sigma_z$ is the chiral (unitary PH) symmetry operator.
Since $\Pi=\mathcal{M}$ it follows that $\mathcal{H}_M=0$,
indicating that there are no mirror symmetric perturbations that can lift the degeneracy of the resulting $N$ MKPs.
This analysis suggests that, in the presence of mirror-line symmetry,
a 1D DIII SC is characterized by an {\it integer}, rather than a $\mathbb{Z}_2$, invariant that determines $N$.

\begin{figure}[t]
\scalebox{0.37}{\includegraphics*{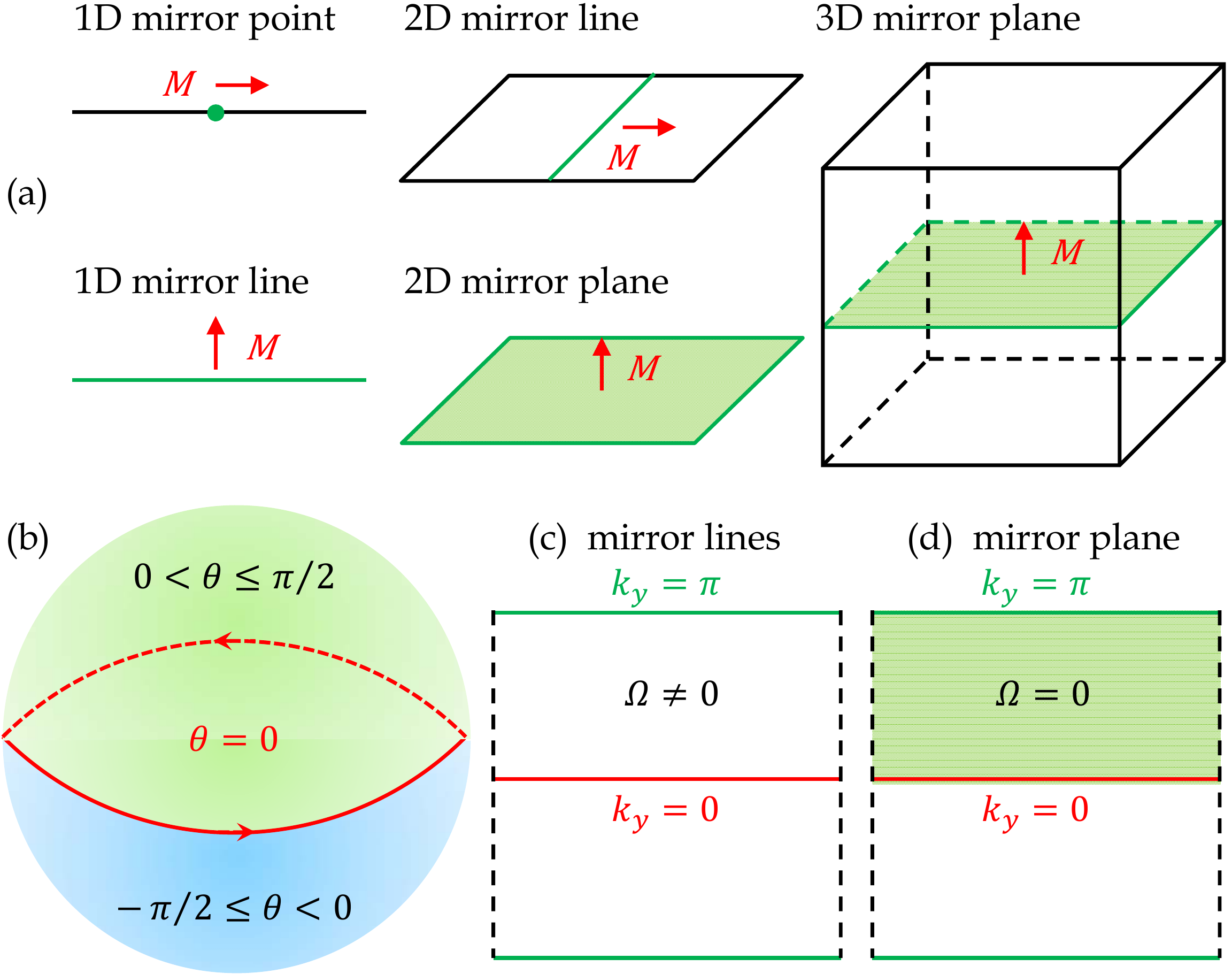}}
\caption{(a) A sketch of various mirror symmetries.
(b) A 1D BZ (red) and an artificial 2D sphere where the chiral symmetry is relaxed.
(c) A 2D BZ with mirror invariant lines (red and green) and nontrivial Berry curvature.
(d) A 2D BZ as a mirror invariant plane in which Berry curvature vanishes.
}\label{fig1}
\end{figure}

{\color{cyan}{\indent{\em Mirror line topological invariant.}}}---
In 1D a DIII SC with a mirror line respects {\it three independent} symmetries:
\begin{eqnarray}
\mathcal{H}_{\phi}\mathcal{M}=\mathcal{M}\mathcal{H}_{\phi}\,,\quad
\mathcal{H}_{\phi}\Theta=\Theta\mathcal{H}_{\bar{\phi}}\,,\quad
\mathcal{H}_{\phi}\Pi=-\Pi\mathcal{H}_{\phi}\,,\quad
\end{eqnarray}
where $\phi\equiv k$, $\bar{\phi}\equiv-\phi$, and $k$ is the momentum.
We are free to choose the pre phase factors of $\Theta$ and $\Xi$ such that $\{\Theta,\Pi\}=0$ and $\Pi^\dagger=\Pi$.
The integer invariant is related to the Berry phase of the negative energy states around the 1D Brillouin zone (BZ).
Since this Berry phase is gauge dependent, however, the gauge needs to be fixed.   This can be accomplished by
introducing a continuous deformation that trivializes the Hamiltonian
by {\it relaxing} the chiral symmetry while {\it keeping} mirror and TR symmetries.
We thus add an artificial dimension $\theta$ ($-\pi/2\leq\theta\leq\pi/2$)
as follows,
\begin{gather}
\mathcal{H}(\theta,\phi)=\mathcal{H}(\phi)\,\cos\theta+\Pi\,\sin\theta\,.
\end{gather}
$\mathcal{H}(\theta,\phi)$ inherits the following symmetry constraints
\begin{subequations}
\begin{eqnarray}
\mathcal{M}^{-1}\mathcal{H}(\theta,\phi)\mathcal{M}&=&\mathcal{H}(\theta,\phi)\,,\label{MS}\\
\Theta^{-1}\mathcal{H}(\theta,\phi)\Theta&=&\mathcal{H}(-\theta,-\phi)\,,\label{TRS}\\
\Pi^{-1}\mathcal{H}(\theta,\phi)\Pi&=&-\mathcal{H}(-\theta,\phi)\,.\label{RCS}
\end{eqnarray}
\end{subequations}
Applying Stokes' theorem, the loop integral of Berry connection~\cite{Niu} along the equator ($\theta=0$) may be written
as the surface integral of Berry curvature~\cite{Niu} $\Omega_{\theta\phi}$ over the north hemisphere ($0\leq\theta\leq\pi/2$), as shown in Fig.~\ref{fig1}(b).
This procedure amounts to choosing a gauge in which the wavefunctions are able to contract into the nonsingular north pole.
The mirror symmetry (\ref{MS}) allows us to label the bands with mirror eigenvalues,
and the total and the mirror Berry phases (in units of $2\pi$) of the valence bands are well defined as
\begin{gather}
\gamma_{t}=\mathcal{C}^{v,+}_{N}+\mathcal{C}^{v,-}_{N}\,,\qquad
\gamma_{m}=\mathcal{C}^{v,+}_{N}-\mathcal{C}^{v,-}_{N}\,.
\end{gather}
Here $\mathcal{C}^{s,i}_{N(S)}$ is the surface integral of
$\Omega^{s,i}_{\theta\phi}$ over the north (south) hemisphere normalized by $2\pi$,
$s=c\,(v)$ denotes the conduction (valence) bands,
$i=+\,(-)$ represents the mirror eigenspace with $i\mathcal{M}=+\,(-)$,
and the sum over unspecified band indices is implicit.

The TR symmetry (\ref{TRS}), the {\it relaxed} chiral symmetry (\ref{RCS}),
and the completeness relation for the energy bands $\sum_{n\in all}|n\rangle\langle n|=1$ respectively lead to
\begin{gather}\label{BCrelation}
\Omega_{\theta\phi}^{s,i}=-\Omega_{\bar{\theta}\bar{\phi}}^{s,\bar{i}}\,,\quad
\Omega_{\theta\phi}^{s,i}=-\Omega_{\bar{\theta}\phi}^{\bar{s},i}\,,\quad
\Omega_{\theta\phi}^{s,i}=-\Omega_{\theta\phi}^{\bar{s},i}\,.
\end{gather}
As a result, $\mathcal{C}^{s,i}_{\alpha}=-\mathcal{C}^{s,\bar{i}}_{\bar{\alpha}}
=-\mathcal{C}^{\bar{s},i}_{\bar{\alpha}}=-\mathcal{C}^{\bar{s},i}_{\alpha}$.
In light of the fact that $\sum_{\alpha=S,N}\mathcal{C}^{s,i}_{\alpha}$ is an integer quantized Chern number, we conclude that
\begin{eqnarray}\label{BTI}
\gamma_t=0\,,\qquad\quad
\gamma_m=\mathbb{Z}\,.
\end{eqnarray}

A qualitative understanding of (\ref{BTI}) is possible. For a mirror line, each mirror subspace respects only chiral symmetry
and thus has an integer topological invariant, {i.e.}, a 1D insulator in class AIII~\cite{Ryu,Kitaev}.
In a gauge where the wavefunctions are contractible to a nonsingular point,
the valence-band Berry phase uniquely characterizes the winding number of its associated Hamiltonian.
It is TR (or PH) symmetry that requires the two invariants belonging to different mirror subspaces opposite to each other.
Consequently, the {\it total} Berry phase must {\it vanish} while the {\it mirror} Berry phase can {\it survive}.

As a consequence, a 1D DIII SC with a mirror line exhibits $|\gamma_m|$ MKPs at the end.
For a 2D DIII SC with a mirror line, there exist {\it two independent integer} numbers of helical Majorana edge states
at the mirror symmetric edge~\cite{Yao}, {i.e.}, $|\gamma_m(0)|$ at $k_y=0$ and $|\gamma_m(\pi)|$ at $k_y=\pi$ along $\hat{x}$.
In the presence of more than one mirror lines, different edges may have different Majorana distributions,
since different mirrors lead to different invariants.
We note that the parity of total mirror Berry phase $(-1)^{\gamma_m}$ in 1D or $(-1)^{\gamma_m(0)\pm\gamma_m(\pi)}$ in 2D
is the $\mathbb{Z}_2$ invariant in class DIII. When $\gamma_m(0)\pm\gamma_m(\pi)$ is odd,
at a mirror {\it asymmetric} edge there also emerges an odd number of helical Majorana edge states protected by TR symmetry.
Since a nontrivial $\mathbb{Z}_2$ index implies a nonzero $\gamma_m$ a DIII topological SC with a mirror line
must also be a topological mirror SC, but not vice versa.
\begin{table}[b!]
\caption{Topological classification of TR invariant SCs and various mirror SCs in zero to three dimensions.}
\newcommand\T{\rule{0pt}{3.ex}}
\newcommand\B{\rule[-1.7ex]{0pt}{0pt}}
\centering
\begin{ruledtabular}
\begin{tabular}{c|cccccc}
       & \multicolumn{2}{c}{Mirror Point} & \multicolumn{2}{c}{Mirror Line} & \multicolumn{2}{c}{Mirror Plane}\\[3pt]
      \hline
      Dim & $D=0$ & $D=1$ & $D=1$ & $D=2$ & $D=2$ & $D=3$ \T\\[3pt]
      \hline
      DIII & $0$ & $\mathbb{Z}_2$ & $\mathbb{Z}_2$ & $\mathbb{Z}_2$ & $\mathbb{Z}_2$ & $\mathbb{Z}$ \T\\[3pt]
      \hline
      Mirror & $0$ & $0$ & $\mathbb{Z}$ & $\mathbb{Z}\times\mathbb{Z}$ & $\mathbb{Z}\times\mathbb{Z}$ & $\mathbb{Z}^4$ \T
\end{tabular}
\end{ruledtabular}
\label{tab:AZ}
\end{table}

{\color{cyan}{\indent{\em Mirror invariant plane.}}}---
We now consider the case for a 2D DIII SC with a mirror plane in which bands can be labeled with mirror eigenvalues.
In each mirror subspace, the completeness relation requires $\Omega_{xy}^{v,i}+\Omega_{xy}^{c,i}=0$
while the chiral symmetry restricts $\Omega_{xy}^{v,i}=\Omega_{xy}^{c,i}$. Therefore, $\Omega_{xy}^{s,i}=0$,
implying both the total and the mirror Chern numbers $\mathcal{C}^{v,+}\pm\mathcal{C}^{v,-}$ are zero.
Applying Stokes' theorem as shown in Fig.~\ref{fig1}(d), the vanishing of the integral of the mirror Berry curvature over a half cylinder
with $-\pi\leq k_x\leq \pi$ and $0\leq k_y\leq \pi$ imposes that
\begin{eqnarray}\label{0=pi}
\gamma_m(0)=\gamma_m(\pi)\,,
\end{eqnarray}
{i.e.}, the mirror Berry phases are the same along the two TR invariant lines $k_y=0$ and $\pi$.
As a result, the number of helical Majorana edge states at $k_y=0$ and $\pi$ is the same.
Furthermore, different edges may have different numbers of helical Majorana edge states,
indicating that the topological classification of mirror-plane SCs is $\mathbb{Z}\times\mathbb{Z}$.
Again, the parity of total mirror Berry phase $(-1)^{\gamma_m(0)\pm\gamma_m(\pi)}$ is the DIII $\mathbb{Z}_2$ index.
Because of (\ref{0=pi}), a 2D DIII SC with a mirror plane is always $\mathbb{Z}_2$ trivial,
however, it can be mirror topological.

Interestingly, any change of $\gamma_m$ from $k_y=0$ to $\pi$ would imply the existence of bulk nodes
which are topologically protected. On the edge parallel to $\hat{y}$,
there would emerge different numbers of helical Majorana edge states across $k_y=0$ and $\pi$,
separated by the projected nodes.
This phenomenon is an analog to the edge state of graphene and the surface Fermi arc of Weyl semimetal~\cite{graphene}.

A mirror plane can also exist in a 3D SC. Assuming $\mathcal{M}_z=-i\sigma_z$ mirror symmetry,
(\ref{0=pi}) can be generalized as
$\gamma_m(k_{\shortparallel}=0,k_z)=\gamma_m(k_{\shortparallel}=\pi,k_z)$,
where $k_z=0$ or $\pi$ is a mirror invariant plane and $k_{\shortparallel}$ denotes $k_x$ or $k_y$.
Clearly, there are {\it four independent} mirror Berry phases for a 3D mirror SC.
At the surface normal to $\hat{k}_{\shortparallel}\times\hat{k}_z$, there distributes $|\gamma_m(0,0)|$ helical Majorana surface states
centered at $(0,0)$ and at $(\pi,0)$, and $|\gamma_m(0,\pi)|$ surface states at $(0,\pi)$ and at $(\pi,\pi)$, respectively.
Different mirror symmetries have difference invariants and
even one mirror invariant plane may have quite different invariants along different directions,
resulting in surface-dependent distributions of Majorana modes.
A 3D SC in class DIII is classified by an integer invariant, however, mirror symmetry requires this integer to be zero.
In the basis where the chiral symmetry operator is $\Pi=\tau_z$,
a 3D DIII SC may be topologically deformed to $\mathcal{H}_{\bm k}=Q^x_{\bm k}\tau_x+Q^y_{\bm k}\tau_y$
which has a flattened eigenvalue spectrum $\mathcal{H}_{\bm k}^2=1$,
and its integer invariant can be understood as~\cite{Ryu,Volovik}
\begin{align}\label{Nw}
N_w=\int\frac{d^3{\bm k}}{24\pi^2}\epsilon^{ijk} {\rm Tr}[(Q^{-1}\partial_i Q)(Q^{-1}\partial_j Q)(Q^{-1}\partial_k Q)]\,,
\end{align}
which is the homotopy of $Q_{\bm k}$$\equiv$$Q_{\bm k}^x-iQ_{\bm k}^y$.
Since $[\mathcal{M}_z,\Pi]$$=$$0$ and $\mathcal{H}_{\bar{k}_z}=\mathcal{M}_z^{-1}\mathcal{H}\mathcal{M}_z$ the integrand of (\ref{Nw})
transforms as a pseudoscalar under mirror operation, {i.e.}, $\rho_w(\mathcal{M}_z{\bm k})=-\rho_w({\bm k})$.
This leads to $N_w=0$, as in the inversion symmetric case.
Therefore, a 3D DIII SC with a mirror plane must be TR trivial, however, it can be mirror topological.

A 0D or 1D DIII SC with a mirror point can be also considered.
At the mirror invariant point, the number of positive and negative energy states is the same because of PH symmetry,
and only the $\mathbb{Z}_2$ parity of the number of negative energy states can be well defined, given that the total charge is not conserved.
TR symmetry further requires an even parity because of Kramers degeneracy.
Therefore, there is no topological classification for a mirror invariant point.
Table~\ref{tab:AZ} summarizes our results in different mirror classes and in different dimensions.

\begin{figure}[t]
\scalebox{0.41}{\includegraphics*{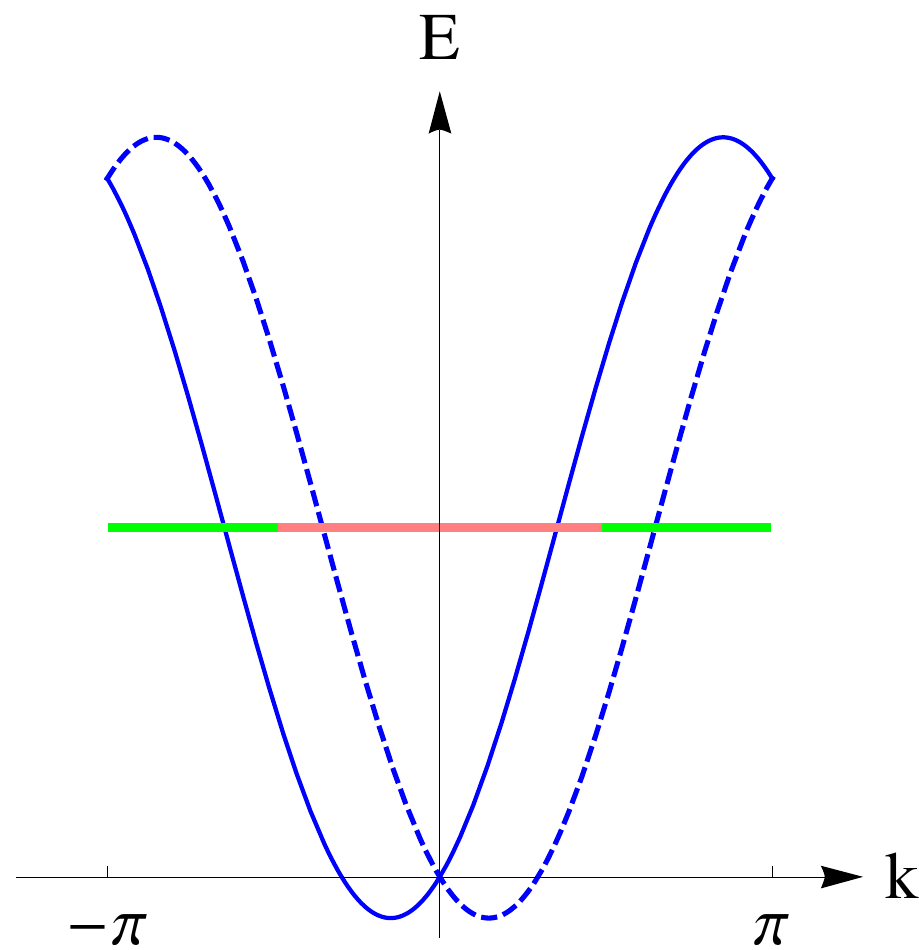}}\quad
\scalebox{0.53}{\includegraphics*{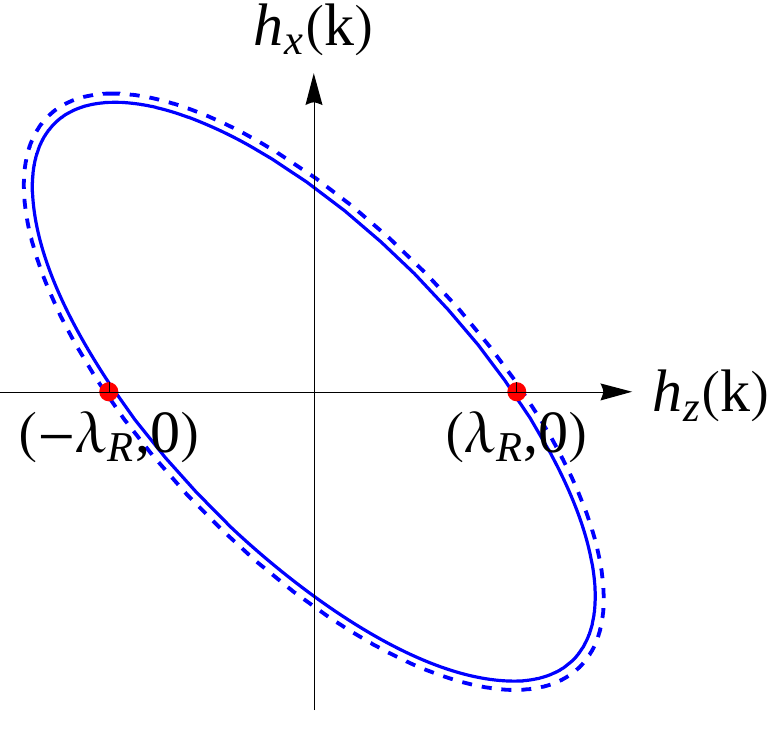}}
\caption{{Left: the two energy bands split by the Rashba spin-orbit coupling;
the pink and green lines depict that the pairing potential switch signs at $k=\pm \pi/2$.
Right: the winding of the vector $\bm{h}^{\pm}(\mu=0)$ as $k$ varies from $0$ to $2\pi$, clockwise (counterclockwise) for $\bm{h}^{+(-)}$.
The solid and dashed lines distinguish the cases for $i\mathcal{M}_z=\pm 1$.
}}\label{fig:berry}
\end{figure}

{\color{cyan}{\indent{\em Representative models.}}}---
The simplest model of topological mirror superconductivity is described by
\begin{eqnarray}
\label{HBdG}
\mathcal{H}=(t\cos k+\lambda_{R}\sin k\,\sigma_z-\mu)\tau_z
+\Delta_1\cos k\,\tau_x\,,
\end{eqnarray}
which may be realized in a Rashba wire that is proximity-coupled to a {\it nodeless} $s_{\pm}$ wave SC~\cite{Zhang-DIII},
{\it e.g.}, an iron-based SC~\cite{Hosono,Mazin}. In this hybrid system, $t$ is the nearest neighbor hopping,
$\mu$ is the chemical potential, and $\lambda_{R}$ is the strength of Rashba spin-orbit coupling.
${\bm \sigma}$ are the Pauli matrices of electron spin while ${\bm \tau}$ are the Pauli matrices in Nambu PH notation.
The order parameter $\Delta_1>0$, induced by the proximity effect, leads to a $s_{\pm}$ wave pairing
that switches sign between $k=0$ and $\pi$.
(\ref{HBdG}) has TR (${\Theta}=\sigma_yK$), PH (${\Xi}=i\sigma_y\tau_yK$), and chiral ($\Pi=\tau_y$) symmetries.
When $|\mu|<\lambda_{R}$, a positive (negative) pairing is induced for the inner (outer) pair of Fermi points,
realizing a 1D ${\mathcal{Z}_2}$ topological SC in class DIII~\cite{Zhang-DIII}.
In addition, there is a mirror ($\mathcal{M}_{z}=-i\sigma_{z}$) symmetry in the bulk and at the ends.
Therefore, this state must also be a topological mirror SC, as predicted earlier.
(\ref{HBdG}) can decompose into $\bm{h}^{\pm}\cdot\bm{\tau}$
with $\pm$ the eigenvalues of $i\mathcal{M}_z$.
As shown in Fig.~\ref{fig:berry}, $(h_z^\pm,h_x^\pm)$ have fixed points $(\pm\lambda_R-\mu,0)$ at $k=\pm \pi/2$.
When $|\mu|<\lambda_{R}$,
the windings of $\bm{h}^{\pm}$ both enclose the origin {\it once} but with {\it opposite} orientations as $k$ varies from $0$ to $2\pi$,
indeed leading to $\gamma_M=1$.

The higher degeneracies of Majorana modes promoted by mirror symmetry are more amazing.
A Majorana quartet~\cite{Zhang-DIII} can be achieved on the $\pi$-junction of two topological SCs modeled by (\ref{HBdG}).
Indeed, a $N$-wire system with each wire described by (\ref{HBdG}) realizes a topological mirror SC with $\gamma_m=N$,
as long as their couplings do not close the bulk gap or break mirror symmetry.
This has been analyzed in the very beginning from the symmetry point of view.
Alternatively, this can be understood by generalizing (\ref{HBdG}) to
$\mathcal{H}_N=\lambda_{N}\sin(N k)\,\sigma_z\,\tau_z+\Delta_N\cos(N k)\,\tau_x$, where we set $\mu=t=0$
and the spin-orbit coupling and pairing become $N$th-neighbor processes as a result of the enlarged unit cell with $N$ sites.
$\mathcal{H}_N$ is the minimal model that describes a topological mirror SC with $\gamma_m=N$.

Furthermore, $\mathcal{H}_N$ can be readily generalized to
\begin{eqnarray}\label{H2line}
\mathcal{H}&=&[\lambda_{N}\sin(N k_x)\,\sigma_z-\lambda_M\sin(M k_z)\sigma_x]\,\tau_z\nonumber\\
&+&[\Delta_0+\Delta_N\cos(N k_x)+\Delta_M\cos(M k_z)]\,\tau_x\,,
\end{eqnarray}
which describes a 2D DIII SC with a $\mathcal{M}_{z}$ mirror line.
When $\Delta_0\neq\pm\Delta_N\pm\Delta_M$, (\ref{H2line}) has a full gap. Along the mirror invariant lines $k_z=0$ and $\pi$,
we obtain
\begin{eqnarray}\label{ML-MBP}
\gamma_m=N\sgn(\lambda_N\Delta_N)\Theta(|\Delta_N|-|\Delta_0+e^{iMk_z}\Delta_M|)\,.\quad
\end{eqnarray}
The $N=M=1$ case can describe a Rashba layer proximity-coupled to a {\it nodeless} $s_{\pm}$ wave SC,
in which one mirror Berry phase vanishes while the other is one when $|\Delta_0|<2|\Delta_1|$.
Because the total mirror Berry phase has an odd parity,
this 2D topological mirror SC is also a DIII topological SC,
consistent with a previous result~\cite{Zhang-DIII}.

Another experimentally {\rm feasible} model that describes a 2D DIII SC with mirror lines is
\begin{eqnarray}\label{H-bilayer}
\mathcal{H}=[\beta k_{\shortparallel}^2-\mu+\alpha(\bm{k}\times\bm{\sigma})_z s_z + m s_x]\tau_z+\Delta s_z\tau_x\,,\quad
\end{eqnarray}
which can be realized in two physical systems.
The first is a thin film~\cite{Zhang-QAH} of topological insulator with a mirror symmetry ($\mathcal{M}_{\shortparallel}$$=$$-i\sigma_{\shortparallel}$), {\it e.g.}, the $\rm Bi_2Se_3$ family.
The $\alpha$-term describes the top and bottom ($s_z$$=$$\pm 1$) surface states that have opposite helicities,
while the {\it smaller} $\beta$-term denotes the surface state curvature which is the same on both surfaces~\cite{Zhang-TISS}.
$m$ is a trivial mass due to the finite size tunneling between the two surfaces.
When $|\Delta|>|m|$, the system is a DIII topological SC since the two surfaces have opposite pairing.
The second system is a Rashba bilayer~\cite{Nagaosa}, {\it e.g.},
the two interfacial 2DEGs ($s_z$$=$$\pm 1$) of $\rm LaAlO_3$-$\rm SrTiO_3$-$\rm LaAlO_3$ sandwich.
The $\beta$-term is the kinetic energy of each 2DEG, while the {\it smaller} $\alpha$-term represents the Rashba spin-orbit coupling~\cite{MacDonald-R} that switches sign on the two interfaces as they are exposed to opposite local electric fields.
$m$ is a small gap at $k=0$ because of interlayer hybridization.
When $|\mu|<|m|$ and $\Delta\neq 0$, only the two outer helical bands are present at Fermi energy
and they acquire opposite pairing, realizing a DIII topological SC.
The odd-parity pairing in (\ref{H-bilayer}) can be engineered via a $\pi$-junction~\cite{Kane-D}
or may be favored by repulsive interactions~\cite{Nagaosa}.
In either case this $\mathbb{Z}_2$ topological SC must also be a topological mirror SC as we have proved.
Near $k_{\shortparallel}=0$ the pairing $\sim s_z\tau_x$ is opposite for different surfaces
while at $k_{\shortparallel}=\infty$ the gap is trivial, indeed leading to $\gamma_m(0)=1$ while $\gamma_m(\pi)=0$.

Finally we consider models for topological mirror-plane SCs.
In the context of $\rm Cu_xBi_2Se_3$, a 3D SC, it is found that the DIII topological state breaks mirror symmetry
whereas the states respecting mirror symmetry are either trivial or nodal~\cite{Fu-DIII,Ando-DIII}.
The nodal states are indeed mirror topological; $\gamma_m$ switches between $0$ and $1$ across the nodes.
These facts are very consistent with our predictions that a SC with a mirror plane must be TR trivial but can be mirror topological.
Our model (\ref{H2line}) also has a mirror plane when $\delta\cos k_y\tau_z$ is added.
For a sufficiently small $\delta$,
the spectrum is fully gapped and all results about (\ref{H2line}) still hold at both $k_y=0$ and $\pi$.

{\color{cyan}{\indent{\em Discussion.}}}---
We have only illustrated the physics of topological mirror superconductivity in simple lattices,
but our theory readily applies to any crystal structure.
A topological mirror SC and its Majorana multiplet are robust
if by average~\cite{average,stat} the disorder respects mirror and TR symmetries.
Besides a mirror fractional Josephson effect,
the MKPs lead to resonant Andreev reflection
producing a pronounced Zero bias conductance peak in tunneling spectroscopy.
Adding a Zeeman field proportional to the mirror operator do not change the invariants,
as long as the reduced gap remains open.
However, the Andreev bound states would be lifted from but still symmetric around zero energy.
As a result, such a field would reduce and split the peak.
In sharp contrast, other Zeeman fields destroy the peak without splitting it.

0D Majorana modes in a 1D topological SC network give hope for fault-tolerant quantum computing~\cite{TQC}.
Manipulating different Majorana modes without coupling or dephasing them is a challenging but necessary task,
which may be solved by adding TR and mirror symmetries.
The TR symmetry likely provides an Anderson's theorem to mitigate the role of disorder in bulk superconductivity;
a mirror symmetry further allows the existence of multiple Majorana modes and more importantly prohibits any coupling among them.
If two topological SCs, with the same (opposite) sign(s) in their mirror Berry phases, respect a common mirror symmetry,
their numbers of MKPs at the linked end are added (subtracted) for two left or two right ends
and subtracted (added) for one left and one right ends.

It is also fascinating to consider other crystalline symmetries.
While inversion is always broken by a boundary, a rotational symmetry can be respected.
It turns out that a rotational symmetry plays two different roles.
Different mirrors may be related by a rotational symmetry and their mirror invariants are hence the same.
A rotational symmetry itself can give rise to topological superconductivity classified by
{\it integer} quantized {\it rotational} Berry phase(s).
For instance, with $\sin k_x\,\sigma_z$ replaced by $\sin k_x\,\sigma_x$ and $k_x\sigma_y-k_y\sigma_x$ replaced by $k_x\sigma_x+k_y\sigma_y$,
our models (\ref{HBdG}) and (\ref{H-bilayer}) respectively describe a 1D and a 2D topological SC protected by
their $\mathcal{C}_2(\hat{x})$ symmetries.
Finally we note that for a class D SC with a mirror plane where there exists a mirror Chern number,
the classification is $\mathcal{Z}$ in 2D and $\mathcal{Z}$$\times$$\mathcal{Z}$ in 3D.
Topological mirror superconductivity will likely open up new horizons for Majorana physics
and even more topological crystalline SCs await to be discovered.

{\color{cyan}\indent{\em Acknowledgements.}}---
It is a pleasure to thank Saad Zaheer and Benjamin Wieder for helpful discussions.
This work has been supported by DARPA grant SPAWAR N66001-11-1-4110.
CLK has been supported by NSF grant DMR 0906175 and a Simons Investigator award from the Simons Foundation.

{\em Note added.}---
After this work was finalized, two complementary and independent studies~\cite{Sato-new,Ryu-new} appeared,
yet their aims and focuses are very different from ours.

\bibliographystyle{apsrev4-1}

\begin{thebibliography}{100}
\bibitem{MKP}
 This is a pair of Majoranas that are TR partners and form a special fermion level.
 States with this level occupied and unoccupied are TR partners and thus TR symmetry acts like a supersymmetry that changes the fermion parity.
\end{thebibliography}

\end{document}